\documentclass{article}

\usepackage{url}
\usepackage{graphicx}
\usepackage{epsfig}
\usepackage{amsmath}

\begin{document}

\title{Memory CODA: introducing memory effects in the Continuous Opinions and Discrete Actions model}

\author{Andr\'e C. R. Martins \\  
	GRIFE -- EACH -- Universidade de S\~ao Paulo,\\
	Rua Arlindo B\'etio, 1000, 03828--000,  S\~ao Paulo, Brazil
}

\maketitle            

\begin{abstract}
The Continuous Opinions and Discrete Actions (CODA) model has been widely used to study the emergence of extremism in social networks. However, this standard model has been shown to generate unrealistic extreme opinions due to the reinforcement among agents. To address this issue, this paper introduces memory effects into the CODA model to explore how the dynamics of opinion formation change. Specifically, each agent is endowed with a memory that stores the previous opinions of its neighbors, which are then utilized to update its own opinion. The paper investigates how incorporating memory affects the strength of choices. We will see that while diminishing the opinion strength, the formation of local domains still causes a significant reinforcement effect. However, unlike the original model, the number of neighbors becomes a relevant variable, suggesting a way to test the results presented in this paper.

\end{abstract}

\section{Introduction}

Extreme opinions can be problematic, depending on what we mean by them \cite{tileaga06a,bafumiherron10a,Martins2022a}. Therefore, understanding how views can become too strong in problems that involve terrorism, destructive polarization, or support for harmful pseudoscience and what we can do about that is a very relevant research field. And part of that investigation has been done using opinion dynamics models~\cite{castellanoetal07,galam12a,latane81a,galametal82,galammoscovici91,sznajd00,deffuantetal00,martins08a,martins12b} to explore how extreme points of view can spread \cite{deffuantetal02a,amblarddeffuant04,galam05,weisbuchetal05,franksetal08a,martins08b,Li2013,Parsegov2017,Amelkin2017} and even how they can emerge from a population of moderate agents and create polarization \cite{DiMaggio1996,Baldassarri2008,Taber2009,Dreyer2019}.

One of the first models where extremism appeared as a consequence of local reinforcement was the Continuous Opinions and Discrete Actions (CODA) model~\cite{martins08a,martins08b}. The basic idea in that model led to several new applications and variations\cite{Deng2013,martins13b,Diao2014,luoaetal14a,Caticha2015,Sobkowicz2018,Lee2018,Garcia2018,Tang2019,Martins2019,Martins2020,LeonMedina2020,Maciel2020,Fang2020,Martins2021,Martins2022a}, including the possibility of integrating other existing models in a general framework ~\cite{martins12b}, something the area does lack~\cite{Boettcher2017,Galam2020}. However, when translated back to probabilities~\cite{martins08a}, the observed opinions go to extreme values that can be considered unrealistic. 

Introducing other effects, such as the influence an agent has over its own neighbors~\cite{martins13c} or the effects of trust~\cite{martins13b} can help decrease the tendency towards unrealistic probability values. Still, it only partially solves the problem. On the other hand, changing the agents' mental models to seek mixed answers instead of certainty \cite{martins16a} can lead to solid but realistic probabilities. However, that happens because agents are no longer looking for the best alternative but a mixture of the two options. When people actually look for one best option, that description of human behavior needs to be corrected. Therefore, even if the model presents a more reasonable outcome, that is not an ideal solution.

To address that problem, I will introduce in this paper memory effects. That is, agents will remember their neighbors' past choices. When those become increasingly repetitive, they will consider it more likely that the neighbors are only keeping an old point of view than actually bringing new information. In the original CODA model, each time an agent observed a neighbor's choice, it considered that as new information, and the agent would update its probabilities accordingly. But that is not a completely realistic description. Once we observe a person who always has the same opinion, learning for the $20^{th}$ time that that person still defends the same choice should be almost no information. Therefore, we will introduce a chance $\mu_k$ in CODA model, where $k$ refers to each of the network-directed links and $\mu_k$ is the probability that agent $j$ (the observed agent in link $k$) choice is only a memory and, therefore, non-informative.

\section{Introducing memory effects in the CODA model}

In the original CODA model, each agent $i$ tries to decide which of two options ($A$ or $B$, or +1 or -1) is the best choice by observing the choices $\sigma_j$ of their neighbors $j$. The full opinion of each agent $i$ is a probability $p_i$ that $A$ is the best choice. Each agent decides its preference simply by checking which option has the most probability. That is $\sigma_i = +1$ if $p>0.5$ and $\sigma_i = -1$ otherwise.

In the more straightforward case, agents also believe their neighbors have a better than random chance $\alpha>0.5$ they will choose the best option, and $\alpha$ does not depend on whether $A$ or $B$ are the actual best option. 

Assuming $i$ observes neighbor $j$ preference and $\sigma_j=+1$, a simple application of Bayes Theorem leads us to the udpate equation
\begin{equation}
	p_i(t+1) = \frac{p_i(t)\alpha}{p_i(t)\alpha+(1- p_i(t))(1-\alpha)}.
\end{equation}
This equation can be simplified by the variable transformation
\begin{equation}\label{eq:nu}
	\nu=\ln(\frac{p}{1-p}). 
\end{equation}
and we have
\[
\nu(t+1) = \nu(t) + C,
\]
where $C=\ln(\frac{\alpha}{1-\alpha})$. We can renormalize that equation so that $C\nu^{*}=\nu$, to make the step equal to 1. Finally, if we calculate the same update rule for the case where $\sigma_j=-1$, we can write both equations as
\begin{equation}\label{eq:CODA}
	\nu^{*}(t+1) = \nu^{*}(t) \pm 1.
\end{equation}
From here on, I will drop the $*$ symbol for simplicity of notation, and $\nu$ should be considered as referring to the renormalized version unless stated otherwise.

\subsection{Adding memory}

To add memory effects, we will consider each agent $i$ remembers the previous opinions of each of their neighbors. For each neighbor $j$, $i$ knows if $j$ had the same choice before and for how many steps $m_{ij}$ agent $j$ has kept that same choice. To implement that, we will use a directed graph  $G = (V, E)$, where $V$ is the set of $n$ nodes corresponding to the $n$ agents in the model, and $E$ is the set of $e$ edges connecting the nodes, representing who is neighbor of whom. In that graph, the memory $m_{ij}$ can be implemented as a property of each directed edge $(i,j)$ in $E$.

Here, we will consider that when agent $i$ observes $j$ choice, $\nu_i$ will be updated by a new rule, where $i$ considers the possibility $j$ has kept the same choice not as new information, but as a possible sign of stubbornness or that $j$ simply did not learn anything new. It is just repeating what it thought before. To introduce that, we assume there is a probability $\mu_{ij}$ that agent $i$ assigns to the possibility $j$ is simply repeating its previous opinion. Obviously, at the beginning of the simulation and when $j$ has just flipped its opinion, we will have  $\mu_{ij}=0$, and the update rule will be identical to the original CODA model. However, for all other cases, we need to understand how 
$\mu_{ij} \neq 0$ changes that rule.

\begin{figure}[h]
	\centering
	\includegraphics[width=0.60\textwidth]{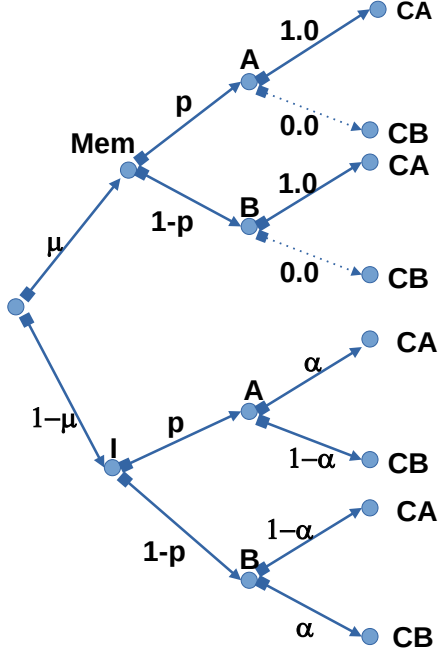}
	\caption{Decision tree for the agent mental model when we include memory effects.}
	\label{fig:tree}
\end{figure} 

Figure~\ref{fig:tree} shows the decision tree that describes the mental model~\cite{Martins2021}, where the indices of  $\mu_{ij}$ and $p_i$ were dropped to make the figure simpler. In the figure, we assume that in the previous observation of $j$, it had chosen $A$, assuming $i$ had observed $j$ preferences before. If it has not, $\mu=0$ and the tree is symmetrical between $A$ and $B$, so defining which choice $+1$ or $-1$ corresponds to $A$ becomes irrelevant. It is easy to notice that if we impose that $\mu=0$, we would use only the lower symmetrical half of the tree, precisely the original CODA model. New effects appear when $i$ observes that $j$ has repeated its previous choice. In that case, we need an initial $0<\mu_{ij}<1$. Here we will assume an initial $M$ probability when $m_{ij}=1$ and  $\mu_{ij}$ is updated together with the internal opinion $p_i$ each time $i$ observes $j$ again. As we will see, the effect of this update is that if $j$ repeats its opinion many times, $i$ will consider, as the simulation proceeds, more and more likely that $j$ has no new information.

In the decision tree, the first level corresponds to an evaluation of whether the agent chose from memory (Mem) or if it should be considered informative (I). The second decision is about which option, $A$ or $B$, is the best. And the third choice is about what agent $j$ actually chooses, depending on the previous nodes, that is, if the neighbor has chosen $A$, that is, $CA$, or $B$, that is, $CB$. That structure corresponds to the probabilities below.

\[
P(CA|A)= \mu + (1-\mu)\alpha,
\] 
and
\[
P(CA|B)= \mu + (1-\mu)(1-\alpha).
\]
Also,
\[
P(CB|A)= (1-\mu) p (1-\alpha)  
\] 
and
\[
P(CB|B)= (1-\mu) (1-p) \alpha  .
\]

Here, we have assumed that $A$ corresponds to the choice +1. When it equals -1, $\alpha$ and $1-\alpha$ should be replaced by each other. This means all equations below should be recalculated. However, the difference in the final equation will just be a sign. We will not repeat all those calculations twice here and just note, for the final equation, that the additive term becomes negative. The equations about $CB$ will not be used, as we assume agent $j$ has chosen $A$ and are shown here just to be complete. Applying Bayes Theorem to obtain an update rule, we get

\[
P(A|CA)=p_i(t+1)=\frac{p_i(t)(\mu + (1-\mu)\alpha)}{N}
\]
and 
\[
P(B|CA)=1-p_i(t+1)=\frac{(1-p_i(t))(\mu + (1-\mu)(1-\alpha))}{N}
\]
where $N$ is a renormalizing constant that gets canceled if we work with the probabilistic odds $o=p/(1-p)$
\begin{equation}
	o(t+1)= o(t)\times \frac{\mu + (1-\mu)\alpha)}{\mu + (1-\mu)(1-\alpha)}.
\end{equation}\label{eq:odds}

In traditional CODA, we would calculate $\nu=\ln(o)$, making Equation~\ref{eq:odds} an additive update rule. However, it is not clear whether we can do that here. Suppose we update $\mu$ as well. In that case, the additive term will change with time, and it could be necessary to keep the probability rule or recalculate probabilities at each implementation step. Therefore, it makes sense to study how  $\mu$ changes with time before deciding on the most efficient way to implement the rule for updating the probabilities.

To obtain an update rule, we need the conditional probabilities related to the memory or information cases, that is,
\[
P(CA|Mem)= 1,
\] 
and
\[
P(CA|I)= p\alpha + (1-p)(1-\alpha).
\]
From that, we get, for the odds associated with $\mu$,
\begin{equation}
	\frac{\mu(t+1)}{1-\mu(t+1)}=\frac{\mu(t)}{1-\mu(t)}\times \frac{1}{ p\alpha + (1-p)(1-\alpha)}.
\end{equation}\label{eq:oddsmu}

Equation~\ref{eq:oddsmu} seems to suffer from the same problem as Equation~\ref{eq:odds}, since it depends on the constantly changing values of $p$. And, if we wanted the final model to be an exact implementation of the Bayesian theorem, we would have to give up using odds or log-odds variables and express everything directly in terms of $p$s and $\mu$s. However, there is something interesting about Equation~\ref{eq:oddsmu}. The term that alters the log-odd depends only on $p\alpha + (1-p)(1-\alpha)$ and that can be seen as an average between $\alpha$ and  $1-\alpha$. If we assume values of $\alpha$ not too far from 0.5, we can use approximate that average as 0.5, and we get the much simpler equation

\begin{equation}
	\frac{\mu(t+1)}{1-\mu(t+1)}=\frac{\mu(t)}{1-\mu(t)}\times 2.
\end{equation}\label{eq:oddsmuapprox}
Equation~\ref{eq:oddsmuapprox} is independent of $p$ and can be used as a simple albeit approximate update rule for the odds. From that, we can be rewrite directly for $\mu(t+1)$ as
\begin{equation}
	\mu(t+1)=\frac{2\mu(t)}{1+\mu(t)}.
\end{equation}\label{eq:muupdate}

It should be noted here that the time $t$ here refers to the number of times neighbor $j$ has repeated its opinion. That is, if we choose the initial $M$ probability for the case where there was no repetition $\mu(t=0)$, Equation~\ref{eq:muupdate} provides us a stable list of values of $\mu(t=0)$ that is valid for any implementation. We can iterate Equations~\ref{eq:muupdate} to calculate the list of $\mu(s)$, where $s$ refers to the number of times $j$ has been observed to have the same choice and use that list during the simulation. That can seriously decrease the number of logarithms the program will calculate, making it more efficient.

That list can be used to generate a list for the second term of the multiplication in Equation~\ref{eq:odds}, and that means we can finally take the logarithm of Equation~\ref{eq:odds} in terms of the steps
\[
S(s)=\ln \left( \frac{\mu(s) + (1-\mu(s))\alpha)}{\mu(s) + (1-\mu(s))(1-\alpha)} \right).
\]
That leads us to a simple rule for updating the log-odds probability $\nu$, after observing agent $j$ has repeated its opinion $s$ times as
\begin{equation}\label{eq:memCODA}
	\nu_i(t+1) = \nu_i(t) + S(s) sign(\nu_j(t)).
\end{equation}
This is a general equation, as we can include the case where the agent has just changed its opinion as the first term in the series, $S(1)=S(\mu(t=0))$, corresponding to only one observation of the same opinion so far. Finally, we can define the renormalized $S(1)\nu^*=\nu$, so that the initial step will be exactly one in $\nu$.

\begin{figure}[ht]
	\centering
	\includegraphics[width=0.75\textwidth]{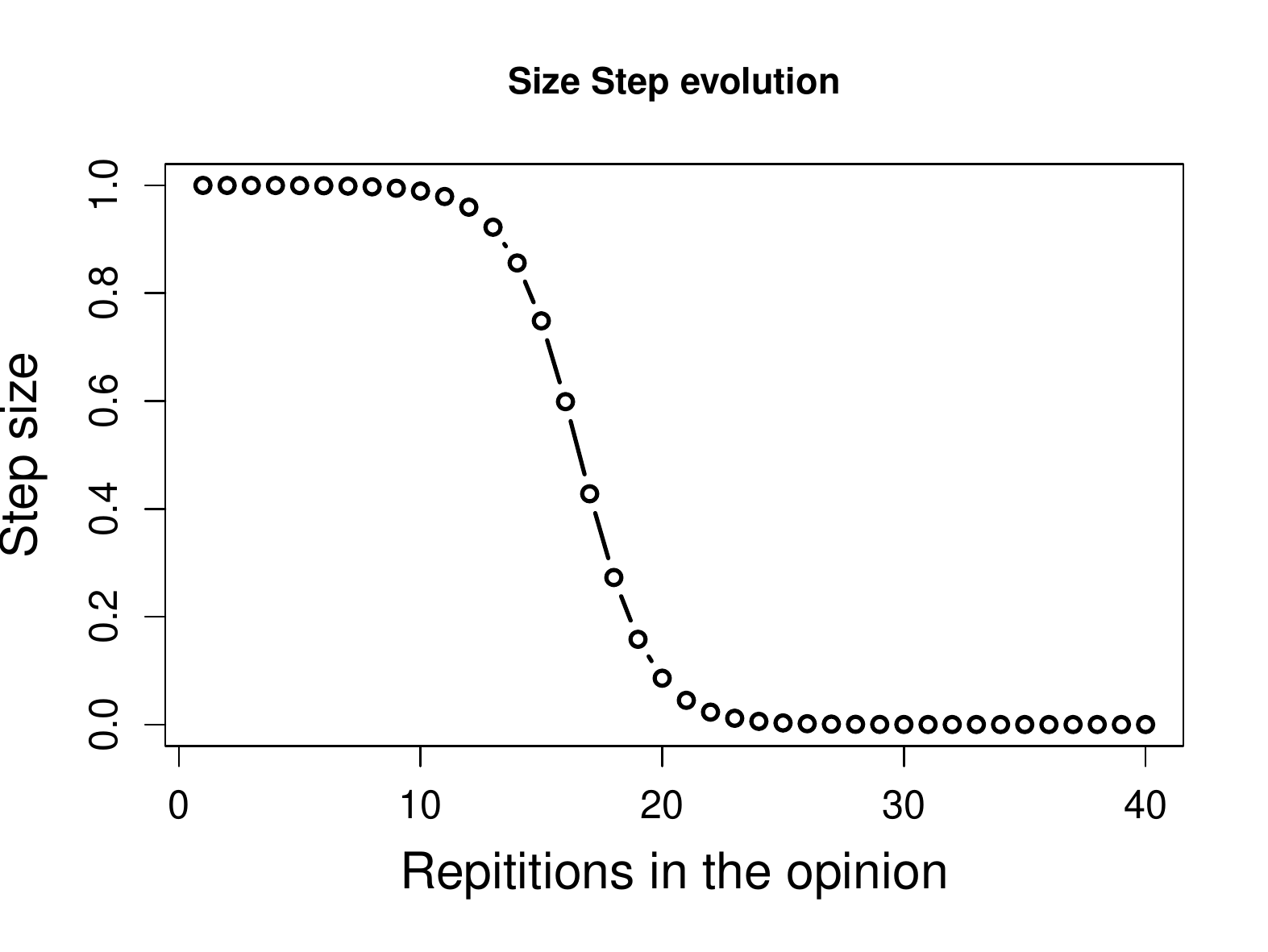}
	\caption{Evolution of the renormalized step sizes as a function of the number of times agent $j$ has repeated its opinion. In this figure, $\mu(t=0)$ was chosen to be atypically small $\mu(t=0)=  0.00001$ to observe the full curve. Typical implementations, starting with a larger $\mu(t=0)$,  normally do not include the left side of the curve.  }
	\label{fig:steps}
\end{figure} 

Figure~\ref{fig:steps} shows how $S$ evolves, starting from a renormalized initial step of 1. The figure corresponds to an unrealistic initial value $\mu(t=0)=  0.00001$, chosen to show the evolution of $S$ if we start from minimal memory effects. More realistic memory effects would basically cause the decrease in step size to happen much sooner.

\section{Results}

The update rule defined by Equation~\ref{eq:memCODA} was implemented using the R software environment \cite{Rsoftware}. The agents were modeled as nodes in a square, two-dimensional, directed lattice without periodic conditions. The simulations shown here correspond to the outcome of $40^2$ agents that observed their neighbors and an average number of 20 observations per agent.

\begin{figure}
	\centering
	\begin{tabular}{c}
		\includegraphics[width=0.85\textwidth]{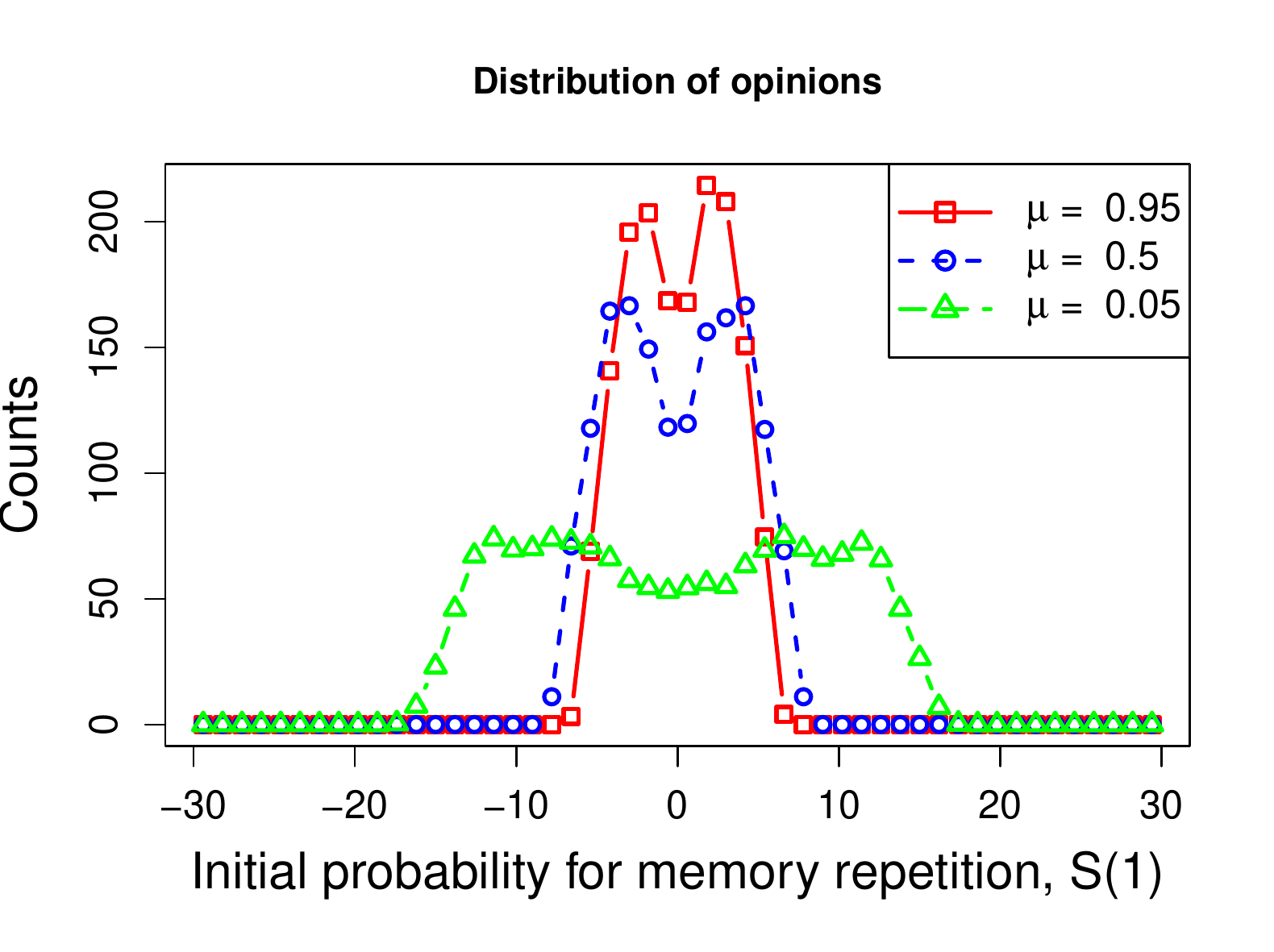}\\
		\includegraphics[width=0.85\textwidth]{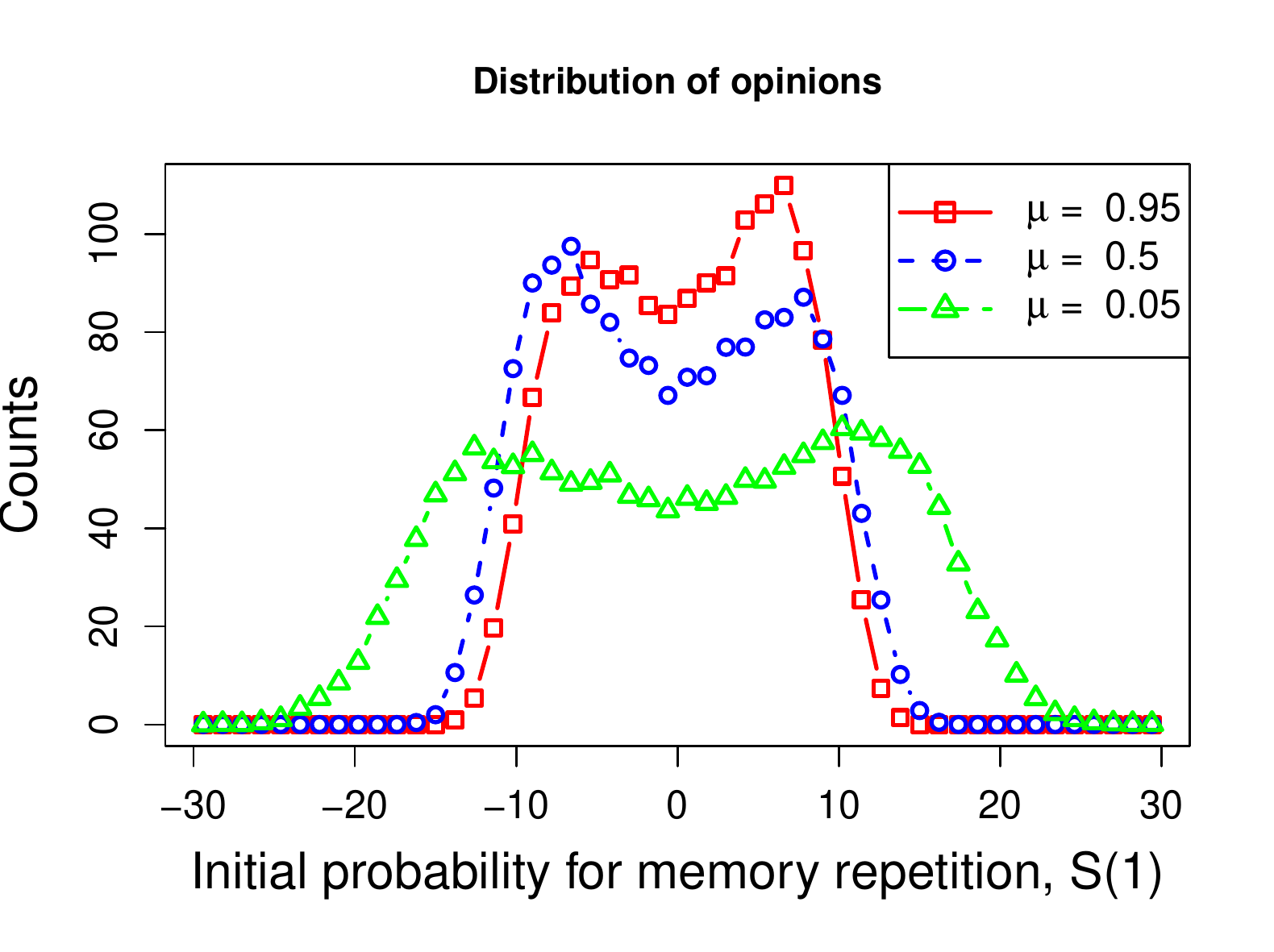}
	\end{tabular}
	\caption{ Distribution of opinions after 20 average interactions per agent  {\it Top panel:} Agents are connected only to their four nearest neighbors. {\it Bottom Panel:} Agents are connected only to their 12 nearest neighbors.
	}\label{fig:opdistr}
\end{figure}

Figure~\ref{fig:opdistr} shows the final average distribution of opinions for different cases. Each curve corresponds to an average of 20 independent cases of the same parameters. In the top panel, agents were connected to their first neighbors. That is, an agent not in the borders would have four neighbors. In the bottom panel, agents were connected up to their second neighbors, totaling 12 connections for those in the center of the graph. We can see clearly in the top panel that, as the initial probability $\mu_0$ assigned to a repetition from memory increases, the final opinions of the agents do become weaker. However, the peaks of extremists still exist. They just happen at less extreme opinion strengths than before. Indeed, in the bottom panel, while a decrease in the strength of opinions is also observed, even when agents consider the first repetition of choice to be caused by memory with a large probability (0.95), we still see the opinions spreading to more significant values and becoming closer to the cases with smaller memory effects.

It is easy to understand why the second case's most extreme opinions are stronger than in the first. Reinforcement of beliefs has two components in the CODA model. One is the continuous repetition, where one observes its neighbors constantly sharing the same choice. This effect is diminished here by the introduction of memory effects. However, opinions also get reinforced by the appearance of domains where almost all agents agree. Inside those neighborhoods, when an agent observed the same neighbor with the same choice, the change in the probabilistic opinion does become weaker. But each distinct neighbor can still contribute, each one starting from the full update. That is, when agents have only four neighbors, the total amount of influence gets much more limited than what we observe when agents have 12 neighbors. The simulations show the opinion results after 20 interactions per agent. Therefore, memory effects are expected to be far less critical when the number of neighbors is close to the average number of interactions. Memory effects still help to decrease the strength of the opinions. Still, their effects become much more significant when agents have a chance to observe their neighbors many times on average.

 \section{Conclusion}
 
The introduction of memory effects in the CODA model, where agents consider that their neighbors might not have any new information and might just be repeating their previous preferences, does decrease the strength of opinions. However, unlike the original version of the CODA model, that decrease depends now significantly on the number of neighbors per agent. That suggests an interesting prediction of the model in this paper, to be verified later. And that is, under regular social interactions, agents who are less connected (fewer friends or a smaller social network) would tend to have weaker opinions than those who are much more connected. Of course, that prediction would not apply to cases of intense indoctrination, where techniques other than just social observation might be relevant.

 	\section{Acknowledgments}
 
 This work was supported by the Funda\c{c}\~ao de Amparo a Pesquisa do Estado de S\~ao Paulo (FAPESP) under grant  2019/26987-2.

\bibliographystyle{unsrt}
\bibliography{biblio}

\begin{thebibliography}{10}

\bibitem{tileaga06a}
Cristian Tileaga.
\newblock Representing the 'other': A discurive analysis of prejudice and moral
  exclusion in talk about romanies.
\newblock {\em Journal of Community \& Applied Social Psychology}, 16:19--41,
  2006.

\bibitem{bafumiherron10a}
Joseph Bafumi and Michael~C. Herron.
\newblock Leapfrog representation and extremism: A study of american voters and
  their members in congress.
\newblock {\em American Political Science Review}, 104(3):519--542, 2010.

\bibitem{Martins2022a}
André~C.R. Martins.
\newblock Extremism definitions in opinion dynamics models.
\newblock {\em Physica A: Statistical Mechanics and its Applications},
  589:126623, 2022.

\bibitem{castellanoetal07}
C.~Castellano, S.~Fortunato, and V.~Loreto.
\newblock Statistical physics of social dynamics.
\newblock {\em Reviews of Modern Physics}, 81:591--646, 2009.

\bibitem{galam12a}
Serge Galam.
\newblock {\em Sociophysics: A Physicist's Modeling of Psycho-political
  Phenomena}.
\newblock Springer, 2012.

\bibitem{latane81a}
B.~Latan\'e.
\newblock The psychology of social impact.
\newblock {\em Am. Psychol.}, 36:343--365, 1981.

\bibitem{galametal82}
S.~Galam, Y.~Gefen, and Y.~Shapir.
\newblock Sociophysics: A new approach of sociological collective behavior:
  Mean-behavior description of a strike.
\newblock {\em J. Math. Sociol.}, 9:1--13, 1982.

\bibitem{galammoscovici91}
S.~Galam and S.~Moscovici.
\newblock Towards a theory of collective phenomena: Consensus and attitude
  changes in groups.
\newblock {\em Eur. J. Soc. Psychol.}, 21:49--74, 1991.

\bibitem{sznajd00}
K.~Sznajd-Weron and J.~Sznajd.
\newblock Opinion evolution in a closed community.
\newblock {\em Int. J. Mod. Phys. C}, 11:1157, 2000.

\bibitem{deffuantetal00}
G.~Deffuant, D.~Neau, F.~Amblard, and G.~Weisbuch.
\newblock Mixing beliefs among interacting agents.
\newblock {\em Adv. Compl. Sys.}, 3:87--98, 2000.

\bibitem{martins08a}
Andr\'e C.~R. Martins.
\newblock Continuous opinions and discrete actions in opinion dynamics
  problems.
\newblock {\em Int. J. of Mod. Phys. C}, 19(4):617--624, 2008.

\bibitem{martins12b}
Andr\'e C.~R. Martins.
\newblock Bayesian updating as basis for opinion dynamics models.
\newblock {\em AIP Conf. Proc.}, 1490:212--221, 2012.

\bibitem{deffuantetal02a}
G.~Deffuant, F.~Amblard, and T.~Weisbuch, G.and~Faure.
\newblock How can extremism prevail? a study based on the relative agreement
  interaction model.
\newblock {\em JASSS-The Journal Of Artificial Societies And Social
  Simulation}, 5(4):1, 2002.

\bibitem{amblarddeffuant04}
F.~Amblard and G.~Deffuant.
\newblock The role of network topology on extremism propagation with the
  relative agreement opinion dynamics.
\newblock {\em Physica A}, 343:725--738, 2004.

\bibitem{galam05}
S.~Galam.
\newblock Heterogeneous beliefs, segregation, and extremism in the making of
  public opinions.
\newblock {\em Physical Review E}, 71:046123, 2005.

\bibitem{weisbuchetal05}
G.~Weisbuch, G.~Deffuant, and F.~Amblard.
\newblock Persuasion dynamics.
\newblock {\em Physica A}, 353:555--575, 2005.

\bibitem{franksetal08a}
Daniel~W. Franks, Jason Noble, Peter Kaufmann, and Sigrid Stagl.
\newblock Extremism propagation in social networks with hubs.
\newblock {\em Adaptive Behavior}, 16(4):264--274, 2008.

\bibitem{martins08b}
Andr\'e C.~R. Martins.
\newblock Mobility and social network effects on extremist opinions.
\newblock {\em Phys. Rev. E}, 78:036104, 2008.

\bibitem{Li2013}
L.~{Li}, A.~{Scaglione}, A.~{Swami}, and Q.~{Zhao}.
\newblock Consensus, polarization and clustering of opinions in social
  networks.
\newblock {\em IEEE Journal on Selected Areas in Communications},
  31(6):1072--1083, June 2013.

\bibitem{Parsegov2017}
S.~E. {Parsegov}, A.~V. {Proskurnikov}, R.~{Tempo}, and N.~E. {Friedkin}.
\newblock Novel multidimensional models of opinion dynamics in social networks.
\newblock {\em IEEE Transactions on Automatic Control}, 62(5):2270--2285, May
  2017.

\bibitem{Amelkin2017}
V.~{Amelkin}, F.~{Bullo}, and A.~K. {Singh}.
\newblock Polar opinion dynamics in social networks.
\newblock {\em IEEE Transactions on Automatic Control}, 62(11):5650--5665, Nov
  2017.

\bibitem{DiMaggio1996}
Paul DiMaggio, John Evans, and Bethany Bryson.
\newblock Have american's social attitudes become more polarized?
\newblock {\em American Journal of Sociology}, 102(3):690--755, 1996.

\bibitem{Baldassarri2008}
Delia Baldassarri and Andrew Gelman.
\newblock Partisans without constraint: Political polarization and trends in
  american public opinion.
\newblock {\em American Journal of Sociology}, 114(2):408--446, 2008.

\bibitem{Taber2009}
Charles~S. Taber, Damon Cann, and Simona Kucsova.
\newblock The motivated processing of political arguments.
\newblock {\em Political Behavior}, 31(2):137--155, Jun 2009.

\bibitem{Dreyer2019}
Philipp Dreyer and Johann Bauer.
\newblock Does voter polarisation induce party extremism? the moderating role
  of abstention.
\newblock {\em West European Politics}, 42(4):824--847, 2019.

\bibitem{Deng2013}
Lei Deng, Yun Liu, and Fei Xiong.
\newblock An opinion diffusion model with clustered early adopters.
\newblock {\em Physica A: Statistical Mechanics and its Applications},
  392(17):3546 -- 3554, 2013.

\bibitem{martins13b}
Andr\'e C.~R. Martins.
\newblock Trust in the coda model: Opinion dynamics and the reliability of
  other agents.
\newblock {\em Physics Letters A}, 377(37):2333--2339, 2013.
\newblock arXiv:1304.3518.

\bibitem{Diao2014}
Su-Meng Diao, Yun Liu, Qing-An Zeng, Gui-Xun Luo, and Fei Xiong.
\newblock A novel opinion dynamics model based on expanded observation ranges
  and individuals’ social influences in social networks.
\newblock {\em Physica A: Statistical Mechanics and its Applications},
  415:220--228, 2014.

\bibitem{luoaetal14a}
Gui-Xun Luo, Yun Liu, Qing-An Zeng, Su-Meng Diao, and Fei Xiong.
\newblock A dynamic evolution model of human opinion as affected by
  advertising.
\newblock {\em Physica A}, 414:254--262, 2014.

\bibitem{Caticha2015}
Nestor Caticha, Jonatas Cesar, and Renato Vicente.
\newblock For whom will the bayesian agents vote?
\newblock {\em Frontiers in Physics}, 3:25, 2015.

\bibitem{Sobkowicz2018}
Pawel Sobkowicz.
\newblock Opinion dynamics model based on cognitive biases of complex agents.
\newblock {\em Journal of Artificial Societies and Social Simulation}, 21(4):8,
  2018.

\bibitem{Lee2018}
Hyun~Keun Lee and Yong~Woon Kim.
\newblock Public opinion by a poll process: model study and bayesian view.
\newblock {\em Journal of Statistical Mechanics: Theory and Experiment}, page
  053402, 2018.

\bibitem{Garcia2018}
Leandro M.~T. Garcia, Ana~V. Diez~Roux, Andr{\'e} C.~R. Martins, Yong Yang, and
  Alex~A. Florindo.
\newblock Exploring the emergence and evolution of population patterns of
  leisure-time physical activity through agent-based modelling.
\newblock {\em International Journal of Behavioral Nutrition and Physical
  Activity}, 15(1):112, Nov 2018.

\bibitem{Tang2019}
Tanzhe Tang and Caspar~G. Chorus.
\newblock Learning opinions by observing actions: Simulation of opinion
  dynamics using an action-opinion inference model.
\newblock {\em Journal of Artificial Societies and Social Simulation}, 22(3):2,
  2019.

\bibitem{Martins2019}
Andr\'e C.~R. Martins.
\newblock Network generation and evolution based on spatial and opinion
  dynamics components.
\newblock {\em International Journal of Modern Physics C}, 2019.

\bibitem{Martins2020}
Andr\'e C.~R. Martins.
\newblock Discrete opinion dynamics with $m$ choices.
\newblock {\em The European Physical Journal B}, 93(1):1, 2020.
\newblock arXiv:1905.10878.

\bibitem{LeonMedina2020}
F.J. Le\'on-Medina, J.~Tena-S\'anchez, and F.J. Miguel.
\newblock Fakers becoming believers: how opinion dynamics are shaped by
  preference falsification, impression management and coherence heuristics.
\newblock {\em Quality and Quantity}, 54:385--412, 2020.

\bibitem{Maciel2020}
Marcelo~V. Maciel and Andr\'e C.~R. Martins.
\newblock Ideologically motivated biases in a multiple issues opinion model.
\newblock {\em Physica A}, page 124293, 2020.
\newblock https://arxiv.org/abs/1908.10450.

\bibitem{Fang2020}
Aili Fang, Kehua Yuan, Jinhua Geng, , and Xinjiang Wei.
\newblock Opinion dynamics with bayesian learning.
\newblock {\em Complexity}, page 8261392, 2020.

\bibitem{Martins2021}
Andre C.~R. Martins.
\newblock Agent mental models and bayesian rules as a tool to create opinion
  dynamics models.
\newblock arXiv:2106.00199, 2021.

\bibitem{Boettcher2017}
Lucas B\"ottcher, Jan Nagler, and Hans~J. Herrmann.
\newblock Critical behaviors in contagion dynamics.
\newblock {\em Physical Review Letters}, 118:088301, 2017.

\bibitem{Galam2020}
Serge Galam and Taksu Cheon.
\newblock Tipping points in opinion dynamics: A universal formula in five
  dimensions.
\newblock {\em Frontiers in Physics}, 8:446, 2020.

\bibitem{martins13c}
Andr\'e C.~R. Martins.
\newblock Discrete opinion models as a limit case of the coda model.
\newblock {\em Physica A}, 395:352--357, 2014.

\bibitem{martins16a}
Andr\'e C.~R. Martins.
\newblock Thou shalt not take sides: Cognition, logic and the need for changing
  how we believe.
\newblock {\em Frontiers in Physics}, 4(7), 2016.

\bibitem{Rsoftware}
{R Development Core Team}.
\newblock {\em R: A Language and Environment for Statistical Computing}.
\newblock R Foundation for Statistical Computing, Vienna, Austria, 2008.
\newblock {ISBN} 3-900051-07-0.

\end{thebibliography}

\end{document}